\documentclass[acmsmall,screen]{acmart}
\usepackage{colortbl}

\usepackage{amsmath,amsfonts}
\usepackage{tikz}
\usepackage{subcaption}
\usepackage{textcomp}

\usepackage{algorithmic}
\usepackage[linesnumbered,ruled,vlined,noend]{algorithm2e}

\usepackage{tabularx}
\usepackage{multirow}  
\usepackage{tablefootnote}
\usepackage{threeparttable}
\usepackage{array}
\usepackage{makecell}

\usepackage[framemethod=tikz]{mdframed}
\usepackage{tcolorbox}
\usepackage{fontawesome5}

\newcommand*\circled[1]{\tikz[baseline=(char.base)]{
            \node[shape=circle,draw,inner sep=1pt] (char) {#1};}}
\AtBeginDocument{%
  }

\setcopyright{cc}
\setcctype{by}
\acmDOI{10.1145/3715767}
\acmYear{2025}
\acmJournal{PACMSE}
\acmVolume{2}
\acmNumber{FSE}
\acmArticle{FSE050}
\acmMonth{7}
\received{2024-09-12}
\received[accepted]{2025-01-14}

\begin{document}

\title{Revolutionizing Newcomers' Onboarding Process in OSS Communities: The Future AI Mentor}

\author{Xin Tan}
\orcid{0000-0003-1099-3336}
\affiliation{%
  \department{State Key Laboratory of Complex \& Critical Software Environment (CCSE), School of Computer Science and Engineering}
  \institution{Beihang University}
  \city{Beijing}
  \country{China}
}
\email{xintan@buaa.edu.cn}

\author{Xiao Long}
\orcid{0009-0005-0809-7470}
\affiliation{%
  \department{CCSE, School of Computer Science and Engineering}
  \institution{Beihang University}
  \city{Beijing}
  \country{China}
}
\email{longxiao@buaa.edu.cn}

\author{Yinghao Zhu}
\orcid{0000-0002-2640-6477}
\affiliation{%
  \department{School of Artificial Intelligence}
  \institution{Beihang University}
  \city{Beijing}
  \country{China}
}
\email{zhuyinghao@buaa.edu.cn}

\author{Lin Shi}
\orcid{0000-0003-1476-7213}
\authornote{Corresponding author.}
\affiliation{%
  \department{School of Software}
  \institution{Beihang University}
  \city{Beijing}
  \country{China}
}
\email{shilin@buaa.edu.cn}

\author{Xiaoli Lian}
\orcid{0000-0002-6100-7068}
\affiliation{%
  \department{CCSE, School of Computer Science and Engineering}
  \institution{Beihang University}
  \city{Beijing}
  \country{China}
}
\email{lianxiaoli@buaa.edu.cn}

\author{Li Zhang}
\orcid{0000-0002-2258-5893}
\affiliation{%
  \department{CCSE, School of Computer Science and Engineering}
  \institution{Beihang University}
  \city{Beijing}
  \country{China}
}
\email{lily@buaa.edu.cn}


\begin{abstract}
  Onboarding newcomers is vital for the sustainability of open-source software (OSS) projects. To lower barriers and increase engagement, OSS projects have dedicated experts who provide guidance for newcomers. However, timely responses are often hindered by experts' busy schedules. The recent rapid advancements of AI in software engineering have brought opportunities to leverage AI as a substitute for expert mentoring. However, the potential role of AI as a comprehensive mentor throughout the entire onboarding process remains unexplored. To identify design strategies of this ``AI mentor'', we applied Design Fiction as a participatory method with 19 OSS newcomers. We investigated their current onboarding experience and elicited 32 design strategies for future AI mentor. Participants envisioned AI mentor being integrated into OSS platforms like GitHub, where it could offer assistance to newcomers, such as ``recommending projects based on personalized requirements'' and ``assessing and categorizing project issues by difficulty''. We also collected participants' perceptions of a prototype, named ``OSSerCopilot'', that implemented the envisioned strategies. They found the interface useful and user-friendly, showing a willingness to use it in the future, which suggests the design strategies are effective. Finally, in order to identify the gaps between our design strategies and current research, we conducted a comprehensive literature review, evaluating the extent of existing research support for this concept. We find that research is relatively scarce in certain areas where newcomers highly anticipate AI mentor assistance, such as ``discovering an interested project''. Our study has the potential to revolutionize the current newcomer-expert mentorship and provides valuable insights for researchers and tool designers aiming to develop and enhance AI mentor systems. 
\end{abstract}



\maketitle

\section{Introduction}

Open source software (OSS) projects are playing an increasingly critical role in the software industry. As a prominent example of open collaboration, OSS development leverages collective wisdom, enabling developers worldwide to collaborate online in creating software products~\cite{levine2014open, scacchi2007free, feller2002understanding}. While this loose organizational structure fosters continuous innovation~\cite{cole2001continuous}, it also leads to a high contributor turnover rate~\cite{zhou2012make}. Thus, a continuous influx of newcomers is critical for the sustainable and healthy development of OSS projects~\cite{zhou2012make}. Simultaneously, newcomers are eager to contribute to OSS as it provides an excellent opportunity to gain real-world software development skills and domain knowledge ~\cite{ye2003toward}.

Despite this mutual interest, newcomers face numerous barriers when attempting to make their first contributions, often resulting in dropouts~\cite{mendez2018open}. The initial challenge is navigating the vast landscape of OSS projects to find one that matches their interests and technical skills, considering the fact that by January 2023, there are over 28 million public repositories on GitHub~\cite{GitHubWikipedia2024}. Even after identifying a project they wish to contribute to, newcomers face significant hurdles during the onboarding process, particularly with projects that have a long development history~\cite{lee2017understanding}. These challenges include finding suitable tasks, learning community processes, configuring the development environment, and then moving on to coding and submitting patches~\cite{crowston2008free}. This process requires not only programming skills but also social skills~\cite{steinmacher2015systematic}.

To facilitate newcomers' engagement, OSS projects have implemented various onboarding strategies, including providing comprehensive README files, detailing contributing guidelines, and labeling ``good first issues'' (GFIs) within their repositories~\cite{kobayakawa2017github, tan2020first}. Although these mechanisms have eased certain barriers to entry, they are not always effective. For instance, Elazhary et al.~\cite{elazhary2019not} find that the contributing guideline structure of many GitHub projects is written to suit maintainers more than newcomers. Even more concerning is that this cognitive mismatch between newcomers and experts often leads to GFIs being unsuitable for newcomers~\cite{tan2020first}.
In addition to these strategies, many OSS communities have devoted experts who are willing to provide technical guidance to newcomers~\cite{balali2018newcomers, tan2023enough}. 
However, newcomers often cannot receive timely responses because experts are extremely busy, which reduces newcomers' willingness to contribute~\cite{steinmacher2013newcomers}. Thus, a critical question arises - what could be an alternative to a community expert instructing newcomers?

Artificial intelligence (AI) might be an excellent solution. The rapid advancement of AI technology has revolutionized coding practices, with tools like \textit{GitHub Copilot} becoming increasingly integrated into daily development workflows~\cite{bird2022taking}. The subsequent release of \textit{GitHub Copilot Workspace} further demonstrates AI's potential to support the entire software development lifecycle~\cite{CopilotWorkspace2024}. AI has also been applied to reduce the burden on experts during the mentoring process, such as using classifiers to help maintainers label GFIs~\cite{xiao2022recommending, huang2021characterizing}. However, these AI technologies are either not specifically designed for OSS newcomers or address only certain aspects of the onboarding process. The potential for AI to serve as a comprehensive mentor throughout the entire onboarding process remains unexplored. Given the rapid advancement of AI, it is crucial to explore how to fully leverage its potential to help newcomers integrate into OSS communities, thus accelerating their onboarding and advancing AI technologies. To this end, this study aims to explore how AI can mentor and support newcomers in onboarding. We propose the following research questions:

\noindent \textbf{\textit{--- RQ1: How are newcomers currently onboarding in OSS communities? Do they use AI to facilitate this process?}} This question aims to reveal the current onboarding process for newcomers and explore the AI techniques that newcomers already used and challenges faced in this process.
    
\noindent \textbf{\textit{--- RQ2: What design strategies for AI mentor can facilitate newcomers' onboarding process?}} This question aims to obtain the design strategies for AI mentor, which has the potential to revolutionize newcomers' onboarding experience and rebuild current expert-newcomer mentorship.

\noindent \textbf{\textit{--- RQ3: How do newcomers perceive the design strategies?}} This question aims to gather feedback on our design strategies to validate our findings.

\noindent \textbf{\textit{--- RQ4: How well does existing research support the design strategies?}} Through this question, we aim to identify the gaps between our design strategies and current research, thereby pointing out potential future directions for AI technology development.

To address these questions, we applied Design Fiction~\cite{blythe2014research} as a participatory design method. This approach emphasizes exploring concepts rather than focusing on specific technologies, making it ideal for emerging fields like AI mentorship, where concrete implementations may not yet exist. We recruited 19 OSS newcomers. Participants were presented with a fictional story about an AI mentor designed to assist newcomers in completing their contributions. Afterward, they discussed their current challenges and provided feedback on the AI mentor's design strategies. Based on these insights, we developed a prototype named ``\textit{OSSerCopilot}'' and gathered further feedback. To assess the current state of knowledge regarding AI mentorship, we also conducted a comprehensive literature review, evaluating the extent of existing research supporting for this concept. 

Our findings reveal that current expert-newcomer mentorship in OSS projects is insufficient, with newcomers often relying on search engines and community resources due to a lack of direct community support. They face difficulties in understanding project structures, identifying appropriate projects, and selecting suitable tasks. To address these challenges, we identified 32 design strategies for an AI mentor through participatory design fiction, aimed at key onboarding steps. For example, participants expected that the AI mentor could recommend projects based on their personalized requirements, such as programming language, technical frameworks, and learning goals. Participants also found the resulting prototype both useful and user-friendly, expressing a willingness to use it in the future, demonstrating the relevance and effectiveness of the proposed strategies. The literature review highlights several underexplored areas where newcomers greatly anticipate AI mentor's assistance, e.g., ``Grasp Project Structure''. Overall, this paper makes the following contributions:

\begin{itemize}
  \item To the best of our knowledge, this is the first comprehensive study to envision the future AI mentor for the whole process of newcomers' onboarding in OSS communities.
  \item The design strategies, derived from newcomers' perspective, offer valuable insights for researchers and tool designers to create and refine AI mentor systems.
  \item The gaps between design strategies and existing literature reveal several urgent yet under-explored aspects of newcomers' needs, highlighting priorities for future research. 
\end{itemize}
 
\section{Related Work}
The sustained influx of newcomers is paramount for the sustainable development of OSS projects~\cite{alexander2002working}. Extensive research has been conducted to identify barriers faced by newcomers, explore associated factors, and devise strategies to facilitate the onboarding process.

Newcomers to OSS projects encounter a multitude of barriers, encompassing both technical and non-technical factors~\cite{balali2018newcomers}. Technical challenges include limitations in domain knowledge and programming proficiency~\cite{lee2017understanding,shibuya2009understanding}, while non-technical obstacles often involve communication difficulties~\cite{tan2019communicate,ducheneaut2005socialization}. Through a qualitative investigation, Shibuya et al.~\cite{shibuya2009understanding} identified that newcomers frequently struggle with task selection due to inadequate documentation, restrictive contributor license agreements, and insufficient technical support. In a comprehensive analysis, Steinmacher et al.~\cite{steinmacher2014preliminary} categorized 58 difficulties into six groups, including newcomers' characteristics and technical hurdles. Expanding on this foundation, Mendez et al.~\cite{mendez2018open} adopted a novel perspective, examining barriers from the standpoint of tools and infrastructures. Their findings revealed that these barriers permeate all six categories of previously established newcomer barrier types. Tan et al.~\cite{tan2023enough} first analyzed the resolution of GFIs, identifying 14 newcomer challenges including understanding, implementing and testing. Several studies~\cite{qiu2019going,padala2020gender} have examined systematic biases that challenge newcomers in OSS communities, with gender bias standing out in these male-dominated spaces~\cite{qiu2019going, padala2020gender}. 

In view of these barriers, various theories and strategies have been proposed to facilitate newcomers' onboarding. Researchers have modeled newcomers' initial behaviors and interactions to identify factors that influence their long-term engagement with OSS communities~\cite{zhou2012make,bao2019large,jensen2011joining}. For instance, Jensen et al.~\cite{jensen2011joining} demonstrated that timely responses—specifically, those received within 48 hours—are positively correlated with sustained participation of newcomers. Steinmacher et al.~\cite{steinmacher2018let} developed guidelines addressing technical skills, contribution processes, and social behavior to support newcomers. Zhou~\cite{zhou2019onboarding} emphasized the need for research into communication, modularization, task division, and the learning processes of experts within OSS projects to enhance newcomer integration. Steinmacher et al.~\cite{steinmacher2016overcoming} introduced FLOSScoach, a portal supporting OSS newcomers. They found it effectively guided newcomers, lowered barriers, and boosted confidence. Dominic et al.~\cite{dominic2020conversational} initially proposed developing a chatbot as a mentor for newcomers but did not thoroughly analyze their needs or validate the requirements. Task recommendation for newcomers has recently emerged as a significant area of interest ~\cite{huang2021characterizing, tan2020first, xiao2022recommending}. Tan et al.~\cite{tan2020first} noted that maintainers often struggle with labeling GFIs, questioning the effectiveness of current practices. Xiao et al.~\cite{xiao2022recommending} and Huang et al.~\cite{huang2021characterizing} implemented the classifiers to automatically label GFIs.
 
In addition to the above online support theories and strategies, mentoring is a prevalent approach within OSS communities~\cite{tomprou2019career}. Considering the sustainability challenges of projects, an increasing number of OSS communities recognize the critical importance of attracting and mentoring newcomers~\cite{fagerholm2014role}. To optimize the mentorship experience, several researchers have explored mentoring process in OSS communities, focusing on its effects~\cite{silva2017students,mota2021google}, barriers ~\cite{balali2018newcomers}, and effective strategies~\cite{balali2018newcomers, balali2020recommending, tan2023understanding}. However, researchers find that formal mentoring is infrequent in OSS projects, primarily due to the significant time and energy demands placed on experts who are often exceedingly busy~\cite{ducheneaut2005socialization}. With the rapid advancement of AI technologies, leveraging an ``AI mentor'' as an alternative to human experts could be a promising solution to assist newcomers. Emerging tools like GitHub Copilot Workspace~\cite{CopilotWorkspace2024} demonstrate the potential of AI-native development environments to address this need. However, since these tools are not specifically designed for OSS newcomers' boarding, their actual needs and how well such tools meet them remain unclear.

Differing from previous studies, our study first investigate the potential role of an AI mentor throughout the entire OSS onboarding process. Using Design Fiction as a participatory method, we identify and develop AI mentor design strategies from the newcomers' perspective, bridging a key knowledge gap in this area.

\vspace{-0.2cm}
\section{Study Design}

\begin{figure}
    \centering
    \includegraphics[width=0.95\linewidth]{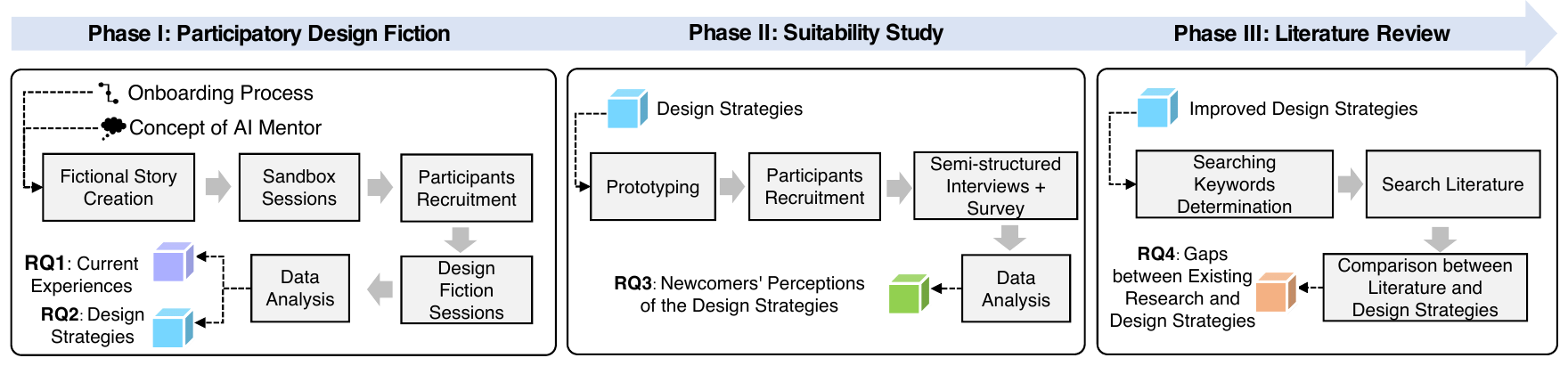}
    \vspace{-0.3cm}
    \caption{Overview of the Research Design}
    \vspace{-0.3cm}
    \label{fig:overview}
\end{figure}
As illustrated in Fig.~\ref{fig:overview}, our study consists of three phases. In phase I, we recruited OSS newcomers and used a design fiction approach to explore their views on potential AI mentor design strategies and their onboarding experiences. In phase II, we developed an AI mentor prototype based on the strategies from phase I and gathered participant feedback. In phase III, we conducted a literature review on how AI supports the onboarding process and compared it to the strategies from the first two phases to identify gaps between existing research and newcomers' expectations. The study was approved by our university's ethics committee.

\vspace{-0.2cm}
\subsection{Phase I: Participatory Design Fiction}
\label{sec:Participatory_Design_Fiction}

We used the Design Fiction method to obtain design strategies for AI mentor 
while simultaneously gaining insights into newcomers' current onboarding experiences. 

The Design Fiction method was first defined by Sterling in 2009~\cite{sterling2009cover} as ``\textit{the deliberate use of diegetic prototypes to suspend disbelief about change}''.
This design practice involves creating tangible and evocative prototypes that envision possible near futures, facilitating the exploration of potential scenarios~\cite{blythe2014research,lindley2016peer}. Because Design Fiction emphasizes conceptual exploration rather than specific technological implementations~\cite{coulton2017design,blythe2016co}, it is well-suited for investigating design strategies in rapidly evolving fields where concrete realizations may not yet exist, such as ``AI mentor''. Given these advantages, Design Fiction has been widely adopted in the Human-Computer Interaction field~\cite{wu2019fair,wong2021using,petsolari2024socio} to elicit information from participants and communicate their insights. For example, Petsolari et al.~\cite{petsolari2024socio} used Design Fiction to explore the role of AI technologies in supporting parenting behaviors and enhancing child well-being across nine future-oriented scenarios.
Recently, the Design Fiction method has also been adopted by some software engineering (SE) researchers~\cite{wessel2022bots, muller2022drinking, barik2016designing, rezwana2023user}. For example, Wessel et al.~\cite{wessel2022bots} obtained a set of design strategies through Design Fiction to enhance how bots present information on pull requests and avoid noise.  Muller et al.~\cite{muller2022drinking} applied Design Fiction to develop conjectures about possible futures of generative AI features and applications.

Past studies~\cite{wessel2022bots, muller2022drinking} utilized Design Fiction to explore and critique potential futures and innovations through narrative and speculative storytelling. Referring to the approach of Wessel et al.~\cite{wessel2022bots}, we applied narratives (i.e., fiction story) and follow-up questions to speculate on the design of future AI mentor for OSS newcomers, shifting focus away from current technical constraints. In the following, we describe how we used the method of Design Fiction in details.

\subsubsection{Fiction Story Creation}
\label{sec:Fiction_Story_Creation}

The story narrates the journey of a newcomer making their first code contributions to an OSS project, guided by an AI mentor. The story description follows the multiple steps of established OSS contribution workflows~\cite{qiu2019signals, crowston2008free}. It begins with the intent to contribute to OSS, followed by the following steps: \circled{1} \textit{discover an interested project}, \circled{2} \textit{understand the project's contribution guidelines}, \circled{3} \textit{grasp the project's structure}, \circled{4} \textit{identify a task (e.g., issues) to work on}, \circled{5} \textit{analyze the issues}, \circled{6} \textit{code a solution}, \circled{7} \textit{test the implementation}, \circled{8} \textit{make the first contribution, such as submit a pull request}, and \circled{9} \textit{modify submitted PR based on the community's feedback}. With the help of AI mentor, the newcomer successfully navigated each step. Not only did they complete their first OSS contribution, but they also gained a deep understanding of the project's technical architecture, became familiar with the contribution process, and significantly improved their skills.

We created a 5-minute animated video to present the fiction story to participants in a standardized way. The protagonist, Alice, is a newcomer eager to make her first OSS code contribution. The narrative is set in the year 2030 considering that the AI mentor is not constrained by current technical limitations. Both the fiction story and the animated video are available in the supplemental material.

\subsubsection{Sandbox Sessions}

We conducted sandbox sessions with a small group of participants to refine the fiction story and overall flow of the session. We invited three undergraduate students who had recently completed a course on OSS development. As newcomers to OSS, they were engaging with OSS for the first time and had recently made, or attempted to make, their initial contributions.

Participants' feedback on the 5-minute animated video helped assess whether the story conveyed its message effectively. We tested the script to ensure it fit within a 90-minute timeframe. After watching the video and answering questions, participants provided insights into the script's flow, leading to minor adjustments. We also analyzed their responses to confirm our research questions could be addressed. The sandbox session data was later discarded.

\subsubsection{Participants Recruitment}

We recruited 19 OSS newcomers, including 10 newcomers from an university OSS development course and 9 newcomers from OSS communities. We conducted the design fiction sessions with 19 participants identified here as P1–P19. Table~\ref{tab:demographics} shows the demographics of our participants. Seventeen participants are male ($\simeq89\%$), while two are female ($\simeq11\%$). Most (14) of them are undergraduate students ($\simeq74\%$), in addition to three postgraduate students ($\simeq16\%$) and two software developers working in IT companies ($\simeq10\%$). Their experience with OSS development (i.e. \#Commits) demonstrates their newcomer status~\cite{tan2020first}: 13 have committed 1 or 2 times ($\simeq68\%$), 6 between 3 and 5 times ($\simeq32\%$). 
As for the experience in software development, 13 ($\simeq68\%$) have less than 2 years of experience, 4 ($\simeq21\%$) between 3 and 5 years, and 2 ($\simeq11\%$) between 5 and 10 years.

\begin{table}[htbp]
\centering
\footnotesize
\caption{Demographics of Participants (\#Years =
Years of experience with software development; \#Commits = Number of commits attempted by newcomers to OSS projects, which may not necessarily be merged)} 
\renewcommand{\arraystretch}{1} 
\begin{tabular}{ccccc|ccccc}
\toprule
ID & Gender & Status & \#Years & \#Commits &  ID & Gender & Status & \#Years & \#Commits\\
\midrule
P1 & Male & undergraduate & 1-2 & 1-2 & P11 & Male & postgraduate & 1-2 & 3-5 \\
P2 & Male & undergraduate & 1-2 & 1-2 & P12 & Male & developer & 3-5 & 3-5 \\
P3 & Male & undergraduate & 3-5 & 3-5 & P13 & Male & postgraduate & 5-10 & 1-2 \\
P4 & Male & undergraduate & 1-2 & 1-2 & P14 & Female  & undergraduate & 1-2 & 1-2 \\
P5 & Male & undergraduate & 3-5 & 3-5 & P15 & Male & developer & 5-10 & 3-5 \\
P6 & Male & undergraduate & 3-5 & 1-2 & P16 & Male & undergraduate & 1-2 & 1-2 \\
P7 & Male & undergraduate & 1-2 & 3-5 & P17 & Male & undergraduate & 1-2 & 1-2 \\
P8 & Male & undergraduate & 1-2 & 1-2 & P18 & Male & undergraduate & 1-2 & 1-2 \\
P9 & Male & undergraduate & 1-2 & 1-2 & P19 & Female  & postgraduate & 1-2 & 1-2 \\
P10 & Male & undergraduate & 1-2 & 1-2 & &&&& \\
\bottomrule
\end{tabular}
\label{tab:demographics}
\end{table}
\vspace{-0.3cm}

\subsubsection{Design Fiction Sessions}
We conducted four same participatory design fiction sessions. Each session includes four to five participants, allowing us to engage everyone in the conversation effectively while keeping the session duration manageable. After conducting the last session, we noticed that there is no new information emerged, which means that we had reached information saturation~\cite{saunders2018saturation}. Thus, we stopped recruiting more participants.
We used a participatory design fiction approach to actively engage participants, allowing real-time questioning during sessions and debriefings. This method gave us the flexibility to explore unexpected information and investigate emerging topics as they arose~\cite{hove2005experiences}. Before each session, participants received a consent form requesting permission for video recording to ensure ethical compliance. We also sent a brief survey with demographic questions to assess their familiarity with OSS development.

Each session began with a brief introduction to the research theme and the Design Fiction approach. We then asked participants about their current onboarding experiences and the AI techniques newcomers used. Afterward, they watched a 5-minute fiction video, followed by a clarification of questions. For each step mentioned in Section~\ref{sec:Fiction_Story_Creation}  (e.g., \circled{1} \textit{discovering an interested project}), we guided discussions around the following three questions: (1) \textit{How did you conduct this step? Did you find it difficult when conducting this step?} (2) \textit{Did you use AI techniques in this step?} (3) \textit{If possible, how would you like AI to help you in this step?}

We asked participants to describe their ideal vision for an AI mentor, unconstrained by current technology. Based on their responses, we followed up with specific questions about the features they mentioned. Following each session, participants were asked to complete a survey designed to gather quantitative data, such as their expectations of AI mentor involvement at various steps of the OSS contribution process. The survey employed Likert scales and sorting questions to measure anticipated levels of AI assistance. The detailed session script and survey questions are available in the supplemental material. Each session was conducted remotely by the first two authors and lasted on average 76 minutes. Participants received \$20 as compensation for their time.

\subsubsection{Data Analysis}

Our data comprises two types: quantitative data, e.g., participants' expectation ratings on each step' assistance provided by AI mentor and answers to Likert-scale questions; and qualitative data, e.g., session audio recordings.

Regarding quantitative data, we undertook descriptive statistical analysis~\cite{bajpai1992descriptive, shull2007guide}. For qualitative data, the session audio recordings (305 minutes) were first transcribed into textual format. We then employed a qualitative analysis approach, utilizing open and axial coding procedures~\cite{strauss1998basics} for the transcripts analysis. The process began with open coding, where the first two authors conducted a preliminary analysis to identify envisioned features of the AI mentor and its environment. This involved an iterative, inductive coding process, building post-formed codes as the analysis progressed. Subsequently, the first two authors engaged in weekly discussions to refine the emergent codes, reaching negotiated agreement~\cite{garrison2006revisiting}. These collaborative sessions aimed to enhance results reliability and mitigate potential bias~\cite{patton2014qualitative}. The analysis was then presented to all co-authors for further discussion. Throughout the process, we applied the constant comparison method~\cite{glaser2017discovery}, continually comparing newly emerging codes with those from previous sessions. Finally, the first two authors performed axial coding to identify relationships between concepts that emerged from the open coding phase.

\subsection{Phase II: Suitability Study}

This phase aims to validate our interpretation of the design strategies from Phase I. Thus, we developed a prototype and collected participants' perceptions.

\subsubsection{Prototyping}

We developed a prototype to gather feedback on the elicited strategies. It simulates the entire newcomer onboarding process, guided by an AI mentor that offers support at each step of the OSS contribution process, as described in Section~\ref{sec:Fiction_Story_Creation}. For example, for \circled{1} \textit{discovering an interested project}, the AI mentor could: (1) \textit{Automatic project recommendation: recommending suitable projects for newcomers based on their historical activities in OSS communities;} (2) \textit{Customized project recommendation: collecting newcomers' personalized requirements for project (e.g. programming language and project domain) through a form, and recommending suitable projects based on these information;} (3) \textit{Helping newcomers analyze whether a project is suitable for contributing.}

In our demonstration, we used the example of a newcomer interested in deep learning projects. By using a GFI in PyTorch project as an example~\cite{PyTorchIssue2024}, we illustrated how our prototype facilitates a smooth onboarding process for OSS newcomers.
We will detail how we mapped the strategies derived from Phase I onto our designed AI mentor interface in Section~\ref{sec:Implemented_Prototype}.

\subsubsection{Interviews}
After completing the prototype, we invited 19 participants to provide feedback via online meetings. This allowed them to comment on our findings~\cite{merriam1998qualitative} and share preferences on the prototype's design elements~\cite{jeffery2017participatory}. All participants, except P5 who was unavailable, joined the sessions. Each session, lasting about 45 minutes, included four to five participants. We demonstrated the prototype, explained the integration of strategies into the interface, and collected feedback. Following each session, participants completed a survey using Likert scales to assess their satisfaction with the AI mentor and the effectiveness of the design strategies.

After transcribing the interviews, the first two authors systematically coded the identified issues and proposed improvements to the prototype. The analysis followed a structured approach similar to participatory design fiction. To ensure reliability, all authors met to discuss and agree on the coding outcomes. Although these interviewees gave valuable feedback, no new suggestions for interface improvements emerged after the fourth interview, indicating information saturation.

\subsubsection{Technology Acceptance Model}
To assess the participants' perception about the designed prototype, we applied the Technology Acceptance Model (TAM)~\cite{davis1989perceived} by conducting a survey following each interview. TAM is a well-established framework for assessing users' perceptions of a technology's usefulness and ease of use, which in turn predicts their acceptance behavior. This model has been widely applied in the SE literature~\cite{steinmacher2016overcoming, mezhuyev2019acceptance,riemenschneider2002explaining,russo2024navigating}. The questions are designed to measure the three main constructs of TAM: perceived usefulness (Ui), ease of use (Ei), and self-predicted future use (Si). We adapted questions from previous research~\cite{davis1989perceived} to fit our AI mentor prototype. A 5-point Likert scale was used, ranging from ``Strongly disagree'' to ``Strongly agree'', with a neutral option. The detailed questions are available in the supplemental material.

\subsection{Phase III: Literature Review}

Existing research has explored automatic tools to assist developers with code contributions. To identify gaps between our design strategies and current research, we conducted a literature review using Google Scholar. We focused on papers published from January 2018 to July 2024 due to the rapid development of AI, particularly generative AI, and to manage manual analysis costs. We included papers from high-quality journals and conferences in SE and AI fields, such as ICSE, ESEC/FSE, ASE, ICPC, SANER, MSR, KBSE, ICSME, PLDI, OOSPLA, TSE, TOSEM, EMSE, JSS, IST, ACL, EMNLP, NAACL, IJCAI, ICLR, NeurIPS, and AAAI.
Then, we established a set of keywords corresponding to each step of the OSS contribution process, as outlined in Section~\ref{sec:Fiction_Story_Creation}. For example, for \circled{1} \textit{Discover an Interested Project}, the keyword set included ``\textit{project/repository recommendation}'', ``\textit{recommend project(s)/repository/repositories}'', and ``\textit{recommend GitHub project(s)/repository/repositories}''. These keywords were based on the AI mentor design strategies from Phases I and II and our understanding of the software development process. We refined them for accuracy with input from a professor specialized in automated software engineering.

Due to the decreasing relevance of later literature and the high cost of manual analysis, we extracted the top 100 relevant papers for each keyword from Google Scholar, removing duplicates based on OSS contribution steps. We reviewed titles and abstracts to ensure the papers focused on automated assistance for specific code contribution steps, excluding irrelevant ones. We analyzed the main contributions, compared them with our design strategies, and identified research gaps. Some papers covered multiple steps because of the overlap between software engineering tasks.  The first two authors discuss the disagreements of the paper and reach an agreement through further paper reading. The complete keywords for each step, along with retrieved and filtered papers, are available in the supplemental material.

\section{Results}

\subsection{RQ1: Current Process of Newcomers' Onboarding and AI Usages}
In this question, we investigated the current onboarding process for newcomers and explored the AI techniques that newcomers already used in the process. Table~\ref{tab:RQ1} details participants' current onboarding process, scenarios of using AI, and challenges for each step.

\begin{table}[]
\centering
\scriptsize
\setlength\tabcolsep{2pt}
\renewcommand{\arraystretch}{1.2}
\caption{Participants' Current Onboarding Experience, Scenarios of Using AI, and Challenges for Each Step}
\label{tab:RQ1}
\renewcommand{\arraystretch}{1.1} 
\begin{tabular}{p{1.3cm}|p{12cm}}  
\toprule
\makecell*[c]{Step} &  \makecell*[c]{Current Onboarding Experience, Scenarios of Using AI, and Challenges}\\
\midrule
\textbf{Discover an Interested Project} & 
\begin{minipage}[t]{12cm}  
        \textbf{Experience:} 1. Contribute to projects currently using or have used (\textbf{5}); 2.Search for projects based on technical skills and interests (\textbf{5}). \\
        \textbf{AI Usages:} 1. Utilize AI to provide a comprehensive overview of project information (\textbf{5}).\\
        \textbf{Challenges:} 1. The restriction of search methods (e.g., only supporting keywords) makes the search process difficult and time-consuming (\textbf{4}); 2. The recommended projects are often large and mature, unsuitable for newcomers (\textbf{4}); \textit{3. The project information provided by AI lacks timeliness (\textbf{3});}  \textit{4. The projects recommended by AI are not actually suitable for newcomers (\textbf{2}).}
\end{minipage} \\ 
 \hline
\textbf{Understand Contribution Guidelines} &
\begin{minipage}[t]{12cm} 
       \textbf{Experience:} 1. Begin with the ``Contribution Process'' section. If issues arise, stop and find relevant information (\textbf{2}).\\
       \textbf{AI Usages:} 1. Utilize AI to comprehend the contributing process (\textbf{4}); 2. Assist in resolving errors encountered during the project deployment process (\textbf{3}); 3. Utilize AI to summarize and translate the contribution guidelines (\textbf{3}).\\
       \textbf{Challenges:} 1. Lengthy project contribution guidelines makes the reading process time-consuming (\textbf{3}).
\end{minipage} \\ 
 \hline
\textbf{Grasp Project's Structure} &
\begin{minipage}[t]{12cm} 
        \textbf{Experience:} 1. Utilize the IDE's code navigation and debugging features to comprehend the project structure (\textbf{6}).\\
       \textbf{AI Usages:} 1. Utilize AI to query programming language syntax and API usage (\textbf{7}); 2. Utilize AI to interpret code (\textbf{6}).\\
       \textbf{Challenges:} \textit{1. It is difficult to comprehend large-scale projects (\textbf{3});} \textit{2. Model has limited input length (\textbf{3});} \textit{3. The performance of AI is highly dependent on the standardization of the project codebase (\textbf{1});} \textit{4. The information returned by AI may be outdated, e.g., an old version of packages (\textbf{1});} \textit{5. Hallucinations errors (\textbf{1}).}
\end{minipage} \\ 
 \hline
\textbf{Identify a Task} &
\begin{minipage}[t]{12cm} 
        \textbf{Experience:} 1. Select issues in interested projects one by one (\textbf{3}); 2. Utilize GFI websites to search for issues (\textbf{2}).\\
       \textbf{AI Usages:} None. \\
       \textbf{Challenges:} 1. Low efficiency (\textbf{3}); 2. Issues labeled as GFIs are unsuitable for newcomers (\textbf{3}).
\end{minipage} \\ 
 \hline
\textbf{Analyze the Issue} &
\begin{minipage}[t]{12cm} 
        \textbf{Experience:} 1. Analyze the code independently (\textbf{3}); 2. Consult community maintainers (\textbf{1}).\\
       \textbf{AI Usages:} 1. Utilize AI to summarize comments from multiple community members on an issue (\textbf{1}); 2. Utilize AI to analyze potential solutions for issues (\textbf{1}). \\
       \textbf{Challenges:} 1. Community feedback is not timely (\textbf{1}).
\end{minipage} \\ 
 \hline
\textbf{Code a Solution} &
\begin{minipage}[t]{12cm} 
        \textbf{Experience:} 1. Write code with the help of search engines, community documentation, and AI (\textbf{1}).\\
       \textbf{AI Usages:} 1. Code complete (\textbf{7}); 2. Debug (\textbf{5}); 3. Develop code frameworks (\textbf{3}); 4. Align code style (\textbf{3}); 5. Write simple or repetitive code (\textbf{2}); 6. Generate unfamiliar code (\textbf{2}); 7. Code translate (\textbf{1}).\\
       \textbf{Challenges:} \textit{1. Hallucinations occur, resulting in the fabrication of non-existent APIs (\textbf{4});} \textit{2. Insufficient support for complex languages, e.g., C++ (\textbf{3});} \textit{3. Generated code may not comply with project package versions (\textbf{1});} 
       \textit{4. Not support cross-file modifications and large-scale projects (\textbf{2});} \textit{5. Response speed is slow (\textbf{1});} \textit{6. The performance is highly dependent on the user's prompting skills (\textbf{1}).}
\end{minipage} \\ 
 \hline
\textbf{Test Implementation} &
\begin{minipage}[t]{12cm} 
        \textbf{Experience:} 1. Conduct testing based on community documentation and historical Pull Requests (\textbf{1}).\\
       \textbf{AI Usages:} 1. Utilize AI to automatically test (\textbf{1}).\\ \textbf{Challenges:} 1. Testing is inadequate (\textbf{1}).
\end{minipage} \\ 
 \hline
\textbf{Make the First Contribution} &
\begin{minipage}[t]{12cm} 
        \textbf{Experience:} 1. Submit code according to contribution guidelines (\textbf{1}); 2. Seek solutions through search engines when encountering problems (\textbf{1}). \\
        \textbf{AI Usages:} 1. Utilize AI to review and improve code (\textbf{1}). \\
        \textbf{Challenges:} None.
\end{minipage} \\ 
 \hline
\textbf{Modify the PR} &
\begin{minipage}[t]{12cm}
        \textbf{Experience:} 1. Communicate with community members to resolve issues (\textbf{1}); 2. Consult community documentation (\textbf{1}). \\
       \textbf{AI Usages:} None. \\
       \textbf{Challenges:} None.
\end{minipage} \\ 
 \hline
\textbf{Others} &
\begin{minipage}[t]{12cm} 
       \textbf{AI Usages:} 1. Draft project documentation (\textbf{8}); 2. Refine communication when interacting with community members (\textbf{2}).
\end{minipage} \\ 
\bottomrule
\end{tabular}
\begin{tablenotes}   
    \footnotesize         
    \item[1] Numbers in parentheses indicate \#participants who mentioned this topic; \item[2] \textit{Italicized text} indicates AI-related challenges.
\end{tablenotes}
\end{table}

\subsubsection{Current Onboarding Process}

In design fiction sessions of Phase I, we asked participants to describe in detail how they performed each step in their current onboarding experience. In general, when newcomers start contributing to OSS projects, they tend to heavily rely on search engines and the thorough examination of existing community resources (e.g., contribution guidelines, historical pull requests, and codebase). Direct guidance or help from the community is less common. For instance, newcomers typically employ two approaches when deciding which project to contribute to. For newcomers who do not have specific projects in mind but are eager to contribute to OSS projects, they often search for projects on platforms like \textit{GitHub} based on their technical skills and interests.  As a participant said, ``\textit{I filtered projects by programming language and sorted them by stars. Then, I checked each project's homepage to see if it matched my interests and had good first issues}'' [P5]. A similar process was followed in step ``\textit{Identify a Task}''. 

In addition to the standard onboarding process, we noticed that the ``Contribution Guidelines'' are a crucial resource for assisting newcomers. They provide essential support and direction, helping newcomers navigate the complexities of contributing to OSS projects. Notably, in the last four steps, multiple participants mentioned that when challenges arise, they frequently rely on the Contribution Guidelines as their go-to source for solutions. For example, ``\textit{...(when reading Contribution Guidelines), I often start by focusing on the contribution process. If I encounter any issues, I then refer to the other sections for additional information}'' [P17].

\subsubsection{Scenarios of Using AI during Current Onboarding Process}

As shown in Table~\ref{tab:RQ1}, AI is most commonly used during the ``\textit{Coding a Solution}'' step, thanks to advancements in code generation technology~\cite{ross2023programmer}. Newcomers rely on AI for code completion, debugging, and aligning code style. AI also helps newcomers' onboarding in two key areas. First, AI serves as a replacement for traditional search engines by helping newcomers quickly retrieve relevant information. For example, newcomers use AI to obtain a comprehensive overview of project information during the ``\textit{Discover an Interested Project}'' step, to resolve errors encountered in project deployment during the ``\textit{Understand Contribution Guidelines}'' step, and to query programming language syntax and API usages during the ``\textit{Grasp Project's Structure}'' step. Second, AI assists newcomers with document-related tasks such as summarizing and translating. For instance, it helps summarize and translate the contribution guidelines during the ``\textit{Grasp Project's Structure}'' step, and it condenses comments from multiple community members on an issue during the ``\textit{Analyze the Issue}'' step. Some newcomers also use AI to draft project documentation and to refine communication when interacting with community members.

\subsubsection{Challenges for Each Step}
\begin{figure}
    \centering
    \includegraphics[width=0.75\linewidth]{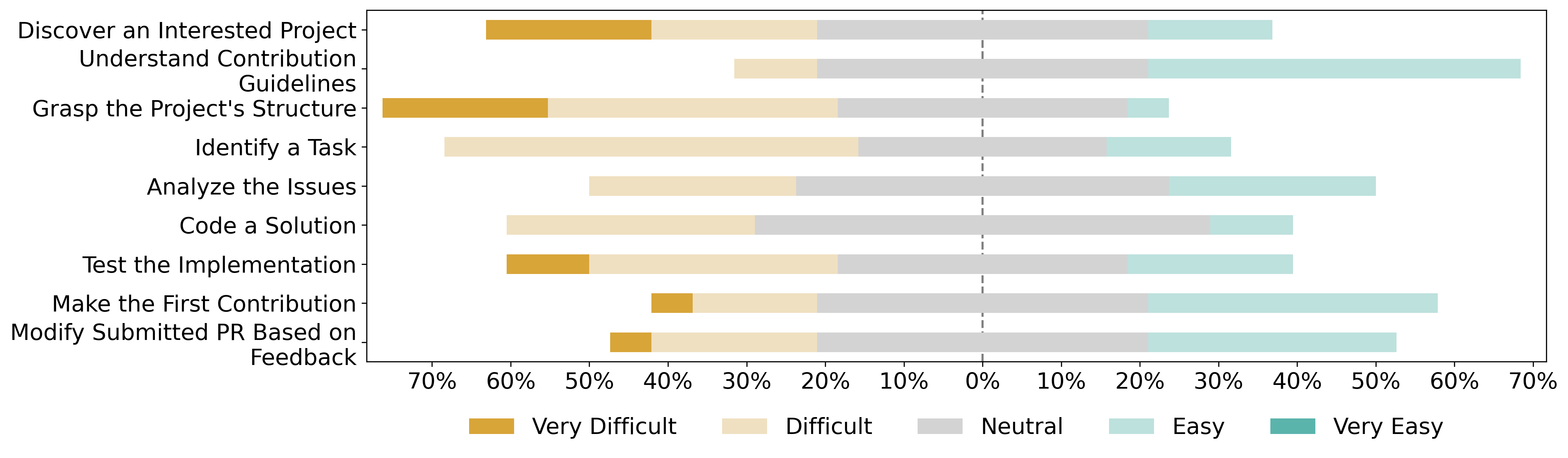}
    \caption{Difficulty Level of Different Steps of Newcomers' Onboarding Process}
    \label{fig:difficulty}
\end{figure}

We surveyed participants after each design fiction session to gather their perceptions of the difficulty of each step in their onboarding process. The results, shown in Fig.~\ref{fig:difficulty}, highlight the three most challenging steps: ``\textit{Grasp the Project's Structure}'', ``\textit{Discover an Interested Project}'', and ``\textit{Identify a Task}'', with over 50\% of participants reporting difficulties. These steps require extensive reading and understanding of the project context, often taking considerable time. As one participant noted, ``\textit{it took me a long time to understand the project's code because the community provided only a brief introduction to its structure}'' [P6].

Interestingly, expect for the three steps (i.e., ``\textit{Understand Contribution Guidelines}'', ``\textit{Make the First Contribution}'', ``\textit{Modify Submitted PR Based on Feedback}''),  for the remaining steps, there were more participants who expressed (very) difficulty than those who held a (very) easy opinion. This overarching trend underscores the intricacies and obstacles that newcomers encounter throughout the OSS onboarding process, emphasizing the need for improved support and guidance to facilitate a smoother transition into the community.

We explored the specific challenges faced by newcomers, as shown in Table~\ref{tab:RQ1}. These challenges fall into two categories: AI-related and AI-unrelated. For AI-unrelated issues, newcomers struggle with the overwhelming amount of information and the lack of beginner-friendly resources. For instance, when ``\textit{Discover an interested Project}'', GitHub's basic search, which relies on keywords and language filters, often returns ``\textit{an overwhelming number of projects}'' [P6], making it hard to narrow down options. P8 noted that \textit{the search function struggles to match their tech stack accurately}. Other challenges include \textit{GFIs not being truly beginner-friendly} [P12] and \textit{slow community feedback} [P15].

AI techniques have made onboarding easier for newcomers, but they have also introduced new challenges. The most frequent AI-related issue was ``\textit{poor performance on large projects}''. Participants noted that AI ``\textit{struggles to comprehend large projects}'' [P3, P18, P19] and is ``\textit{unable to support cross-file modifications and large-scale projects}'' [P7, P12]. Timeliness was another concern, with some saying the ``\textit{AI-provided project overview is outdated}'' [P2, P3, P19] and the ``\textit{AI-generated code doesn't follow the latest package conventions}'' [P1]. Other issues included ``\textit{limited input length}'' and ``\textit{insufficient support for complex languages like C++}''.

\vspace{0.2cm}
\begin{mdframed}
[linecolor = gray!100,linewidth = 3pt,
innerleftmargin = 3pt, topline=false, rightline=false, bottomline=false, leftline=true, innerrightmargin = 3pt,innertopmargin = 3pt, innerbottommargin = 3pt,backgroundcolor = gray!30]
\textbf{Summary for RQ1:} 
Our study reveals that newcomers' onboarding in OSS projects mainly relies on search engines and community resources, with little direct community help. They face significant challenges in understanding project structures, finding suitable projects, and identifying tasks. Although AI tools help with coding and retrieving information, they also present issues like struggling with large projects and outdated content. Overall, better onboarding support and guidance are needed to help newcomers integrate into OSS communities more smoothly.
\end{mdframed}

\subsection{RQ2: Design Strategies for the AI Mentor}

In RQ2, we aim to obtain the design strategies for AI mentor. These strategies have potential to revolutionize the onboarding experience for newcomers and rebuild current expert-newcomer mentorship. We also investigate the extent to which newcomers expect the AI mentor to provide assistance and guidance at different steps, in order to evaluate the priority of these expectations.

\subsubsection{Ranking of Expectations for AI Mentor Assistance at Different Steps}
\label{sec:Expectations_Ranking_at_Different_Steps}

\begin{figure}[htbp]
    \centering
    \begin{minipage}{0.45\textwidth}
        \centering
        \includegraphics[width=\textwidth]{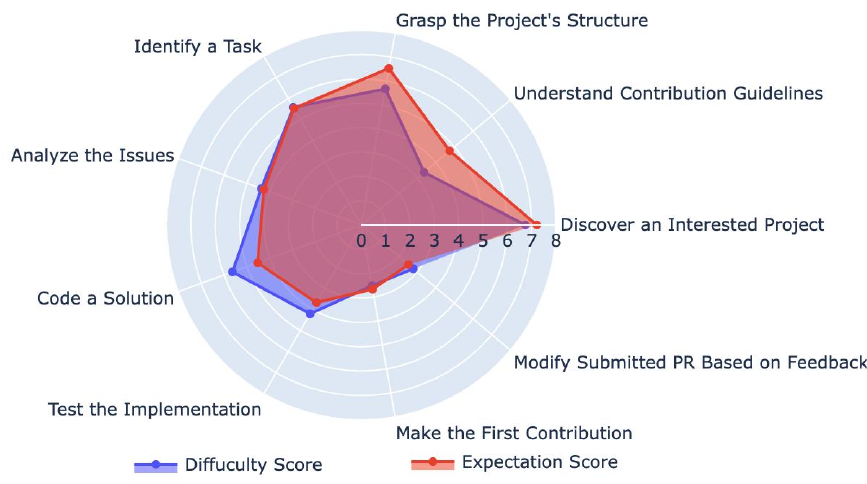}
        \caption{Participants' Perceived Difficulty VS. Expectations for AI Mentor's Assistance}
        \label{fig:difficulty_expectation}
    \end{minipage}
    \hfill
    \begin{minipage}{0.45\textwidth}
        \centering
        \includegraphics[width=\textwidth]{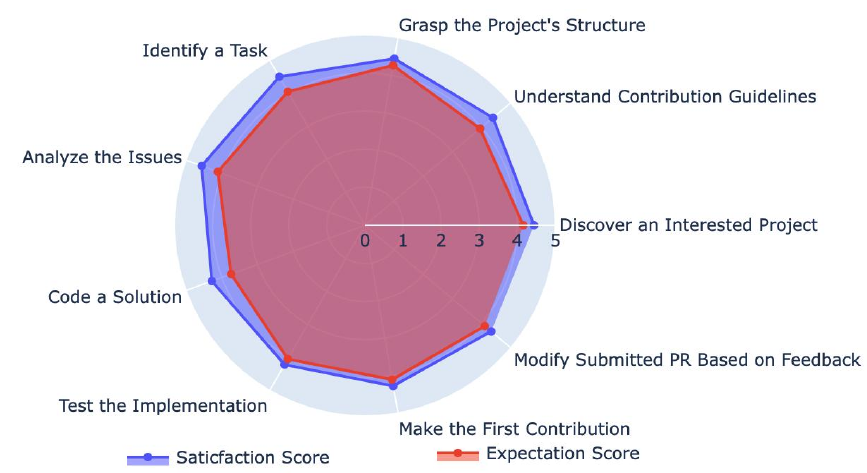}
        \caption{Participants' Expectations for AI Mentor's Assistance VS. Satisfaction with Our Prototype}
         \label{fig:expectation_implementation}
    \end{minipage}
\end{figure}

In the post-session survey, we included two ranking questions to assess the priority of participants' expectations for AI mentor assistance and their perceived difficulties at different steps. We assigned a weighted score to each step based on participants' rankings, with the top-ranked step receiving 9 points and the last-ranked step receiving 1 point (for 9 steps in total). We then calculated the average ranking score for each step to quantitatively assess its importance or difficulty. We compared the average scores for expectations and perceived difficulties for each step, as shown in Fig.~\ref{fig:difficulty_expectation}.

The three steps where participants most desired AI assistance were ``\textit{Discover an Interested Project}'' (average score: 7.21), ``\textit{Grasp the Project's Structure}'' (6.53), and ``\textit{Identify a Task}'' (5.53). These steps were also seen as the most difficult, indicating a strong need for better support. Interestingly, the link between difficulty and the desire for AI help was not always consistent. For example, ``\textit{Understand Contribution Guidelines}'' was rated less difficult (3.37, ranked \#7) but still had a high demand for AI help (4.74, ranked \#4), showing a need for AI to organize, summarize, and answer questions. On the other hand, despite the difficulty of ``\textit{Code a Solution}'' (5.58, ranked \#4), participants did not expect much AI help, with one noting that ``\textit{too much AI assistance could hinder newcomer growth}'' [P2].

\subsubsection{Specific Expectations for AI Mentor Assistance at Different Steps}

\begin{table}[]
\centering
\scriptsize
\setlength\tabcolsep{2pt}
\renewcommand{\arraystretch}{1.2}
\caption{Envisioned Design Strategies for AI Mentor for Each Step}
\label{tab:design_strategies}
\renewcommand{\arraystretch}{1.1} 
\begin{tabular}{p{1.3cm}|p{12cm}}  
\toprule
\makecell*[c]{Step} &  \makecell*[c]{Design Strategies for AI Mentor}\\
\midrule
\textbf{Discover an Interested Project} & 
\begin{minipage}[t]{12cm}  
        1. Recommend projects based on specific requirements provided by newcomers (e.g., programming language, technical frameworks, and learning goals) (\textbf{8}); 
        2. Recommend projects based on newcomers' activity history in OSS community (\textbf{6}); 
        3. Analyze project-related data (e.g., project activity, and the quantity and quality of the project's GFIs) to assist newcomers in assessing whether the project is suitable for contribution.(\textbf{4});
        4. Provide specific details, such as reasons for recommendation, activity level of the project, and the set of GFIs, when recommending projects (\textbf{4}); 
        5. 
        Avoid recommending projects to developers who are actively contributing, while suggesting projects to those frequently browsing GFIs to minimize unnecessary interruptions (\textbf{2}).
        
\end{minipage} \\ 
 \hline
\textbf{Understand Contribution Guidelines} &
\begin{minipage}[t]{12cm} 
       1. Summarize, analyze, and translate contribution guidelines for newcomers (\textbf{6}); 
       2. Offer customized guidance for newcomers on local project deployment based on their development environment (\textbf{5});
       3. Provide guidance during newcomers' contribution process, helping them understand their current stage (\textbf{4}); 
       4. Provide a knowledge-based Q\&A system for the contribution guidelines, allowing newcomers to ask questions as needed to improve reading efficiency and prevent information overload (\textbf{2});
       5. Provide guidance to newcomers on using the development tools specified by the project (\textbf{2}).
       
\end{minipage} \\ 
 \hline
\textbf{Grasp Project's Structure} &
\begin{minipage}[t]{12cm} 
        1. Provide project code explanations at the module, directory, file, or function level (\textbf{6}); 2. The granularity of code analysis should be determined by the newcomer and can be expanded progressively (\textbf{3}).
\end{minipage} \\ 
 \hline
\textbf{Identify a Task} &
\begin{minipage}[t]{12cm} 
        1. Conduct a comprehensive assessment of the difficulty of project issues, then sort or categorize them by difficulty (\textbf{8});
        2. Different projects may use various labels to indicate ``GFI''. Recommend suitable issues for newcomers to contribute to, even if some issues are not labeled as ``GFI'' (\textbf{4});
        3. Provide specific recommending 
        reasons 
        (e.g., how the issue matches the newcomer's needs and profile) (\textbf{4}); 
        4. Indicate the ``significance'' of the issue when recommending,  including its impact on the community or project and the learning opportunities for newcomers (\textbf{2}).
\end{minipage} \\ 
 \hline
\textbf{Analyze the Issue} &
\begin{minipage}[t]{12cm} 
        1. Navigate and locate relevant code of the issue (\textbf{4});
        2. Recommend similar/related issues (\textbf{3});
        3. Reproduce the issue (for bug fix issues) (\textbf{3});
        4. Analyze the steps for resolving the issue(\textbf{2});
        5. Analyze similar existing features for reference when recommending new feature issues. (\textbf{1}).
\end{minipage} \\ 
 \hline
\textbf{Code a Solution} &
\begin{minipage}[t]{12cm} 
        1. Align the code style (\textbf{5});
        2. Explain API usage in the context of project-specific characteristics (\textbf{5});
        3. Help debugging in the context of project-specific characteristics (\textbf{3}). 
\end{minipage} \\ 
 \hline
\textbf{Test Implementation} &
\begin{minipage}[t]{12cm}
        1. Generate unit tests based on the community's testing requirements (\textbf{4});2. Generate test cases for higher test coverage (\textbf{4}). 
\end{minipage} \\ 
 \hline
\textbf{Make the First Contribution} &
\begin{minipage}[t]{12cm} 
        1. Simulate the PR submission process and offer preliminary code modification suggestions before formal submission (\textbf{5});
        2. Adjust newcomer-written code to ensure it complies with community coding standards (\textbf{4});
        3. Revise the PR description or commit messages to facilitate better communication with the community (\textbf{2}). 
\end{minipage} \\ 
 \hline
\textbf{Modify the PR based on Feedback} &
\begin{minipage}[t]{12cm} 
        1. Summarize the reviewers' feedback and provide guidance on how to approach the necessary modifications (\textbf{3});
        2. Analyze the PR's current status in the automated merging process and provide the possible error reasons if the automation halts (\textbf{2}).
\end{minipage} \\ 
 \hline
\textbf{Deployment Modes of the AI Mentor} &
\begin{minipage}[t]{12cm} 
       \textbf{Integrate the AI Mentor with Existing OSS Development Platforms (12): }  \\
       1. Do not require to switch between environments, which avoid disrupting the workflow and improve potential efficiency loss (\textbf{4});
       2. AI mentor can access information more extensively and conveniently (\textbf{3}); 3. The high activity levels on existing OSS platforms create an advantageous environment for the promotion and development of AI mentor. (\textbf{1}).  \\ 
       \textbf{Not Integrate AI Mentor with Existing OSS Development Platforms (3): }  \\
       1. AI mentor can access additional information beyond what is available within the OSS community or platform, such as the newcomer's local code files (\textbf{2}); 2. Prevent information overload (\textbf{2}). 
\end{minipage} \\ 
\bottomrule
\end{tabular}
\begin{tablenotes}   
    \footnotesize         
    \item[1] Numbers in parentheses indicate \#participants who mentioned this design strategy.
\end{tablenotes}
\end{table}

Participants discussed design strategies for the AI mentor in participatory design fiction sessions (refer to Section~\ref{sec:Participatory_Design_Fiction}). After the analysis of the session transcripts, we found 32 design strategies, as outlined in Table~\ref{tab:design_strategies}. Most of these strategies are self-explanatory. We provide further explanation only for those strategies that are either mentioned more frequently or are less straightforward to understand.

\textbf{Discover an Interested Project.} 
Participants identified five strategies for how the AI mentor could help in finding suitable projects. The most mentioned strategy, cited by nearly half of the participants, was ``\textit{recommending projects based on newcomers' specific requirements}''. This aligns with the challenge of ``\textit{Restriction of Search Methods}'' noted in RQ1.  Participants wanted AI recommendations based on more personalized criteria beyond current filters like programming languages. These personalized factors include ``\textit{project technical stacks}'' [P1, P8, P18], ``\textit{project domain}'' [P4, P15], and even ``\textit{newcomers' specific learning goals}'' [P4, P13, P16]. One participant suggested AI could \textit{gather tailored project requirements via a customized form} [P8]. Participants also underscored the importance of ``\textit{offering detailed reasons for recommendations}'' (4) and emphasized the need to consider the timing of these recommendations to ``\textit{minimize unnecessary interruptions}'' (2).

\textbf{Understand Contribution Guidelines.} In this step, participants expected AI mentor to enhance their reading and learning efficiency while preventing information overload. Suggested strategies include ``\textit{summarizing, analyzing, and translating contribution guidelines}'' (6), as well as ``\textit{providing a Q\&A system based on contribution guidelines}'' (2). The second and fourth strategies outlined in this step suggest that AI mentor should provide more personalized guidance on aspects related to development environment setup, including ``\textit{local project deployment}'' (5) and ``\textit{usage of development tools}'' (2). Furthermore, some participants advocate for AI mentor to provide ``\textit{continuous guidance throughout the entire contribution process, helping them understand their current stage}'' (4).

\textbf{Grasp Project's Structure.} Participants suggested that AI mentor should ``\textit{provide more detailed support in understanding project structure} (6) and \textit{allow them to specify the desired level of code analysis granularity} (3)''. As a participant said, ``\textit{the analysis can start at the directory level, letting developers dive deeper into files and even explore specific functions. In the end, AI mentor should display a tree diagram that visually maps out the project's entire structure}'' [P11].

\textbf{Identify a Task.} The first two design strategies in this category further underscore the limitations inherent in the current GFIs mechanism~\cite{tan2020first}. AI mentor should be capable of ``\textit{conducing a more comprehensive difficulty assessment of project issues, and categorizing and ranking them based on difficulty}'' (8), thereby enabling newcomers of varying skill levels to select issues that align with their abilities. Moreover, given the diversity and inconsistency in GFI label naming across different projects, AI mentor should ``\textit{recommend suitable issues for newcomers to contribute to, even if some issues are not labeled as GFI}'' (4).

\textbf{Analyze the Issue.} At this step, participants proposed different design strategies tailored to various types of issues. For example, for ``bug fix issues'', participants expected AI mentor to provide support during the process of reproducing bugs, which could ``\textit{greatly facilitate the resolution of such issues}'' [P7]. Participants also hoped that the AI mentor could guide newcomers in ``\textit{locating code related to the issue}'' (4), or ``\textit{recommending similar/related issues for reference and learning}'' (3).

\textbf{Code a Solution.} As mentioned in Section~\ref{sec:Expectations_Ranking_at_Different_Steps}, newcomers did not wish for the AI to provide excessive code assistance at this step. Therefore, the strategies proposed by participants primarily focus on areas such as ``\textit{aligning code style}'' (5) and ``\textit{explaining API usage}'' (5) — tasks that are time-consuming but do not require much creative effort.

\textbf{Test the Implementation.} This step involves two design strategies, including ``\textit{generating unit tests based on the community’s testing requirements}'' and ``\textit{generating test cases for higher test coverage}''. Both strategies aim to improve the accuracy and coverage of testing. Participant P8 stated, ``\textit{testing is a challenging process. Many times, I'm not sure how to write tests for my code. I hope AI mentor can help me write tests based on community requirements so that I can learn from it}''.

\textbf{Make the first Contribution.} At this step, participants proposed an intriguing strategy - ``\textit{simulating the PR submission process}'' (5), which aims to help newcomers familiarize themselves with the PR submission workflow and get some preliminary code modification suggestions before formal submission. Participants also proposed two design strategies for AI mentor to assist in modifying PR content, thereby ensuring closer alignment with community standards (e.g., coding standards, PR description standards, and commit message standards).

\textbf{Modify the PR based on Feedback.} In addition to ``\textit{providing modification guidance based on reviewer's feedback}'' (3), participants also expected AI mentor to ``\textit{analyze the PR's current status in the automated merging process and provide the possible error reasons if the automation halts}'' (2). As a participant said, ``\textit{as a newcomer, I often encounter various errors during the automated PR merging process, but I'm not familiar with the reasons behind them. I hope AI mentor can indicate the current status of the PR, guide me in analyzing the causes of these errors, thereby making me more familiar with the PR merging process}'' [P16].

\textbf{Deployment Modes of AI Mentor.} In the final step of the design fiction session, we discussed the ``deployment modes of AI mentor''. Most participants (12) preferred \textbf{integrating AI mentor with existing OSS development platforms}, believing this would ``\textit{avoid disrupting the workflow and reduce potential efficiency loss}'' (4), ``\textit{facilitate access to existing platform information}'' (3), and ``\textit{leverage high platform activity to enhance AI mentor}'' (1). However, some participants (3) raised concerns that integration might ``\textit{limit access to information beyond the platform (e.g., local code files)}'' (2) and ``\textit{lead to information overload on platforms}'' (2).

\vspace{0.2cm}
\begin{mdframed}
[linecolor = gray!100,linewidth = 3pt,
innerleftmargin = 3pt, topline=false, rightline=false, bottomline=false, leftline=true, innerrightmargin = 3pt,innertopmargin = 3pt, innerbottommargin = 3pt,backgroundcolor = gray!30]
\textbf{Summary for RQ2:} 
Through participatory design fiction, we elicited 32 design strategies for future AI mentor throughout newcomers' entire onboarding process. Participants most strongly desired AI mentor's assistance in the steps ``\textit{Discover an Interested Project}'', ``\textit{Grasp the Project's Structure}'', and ``\textit{Identify a Task}''. Notably, the relationship between the perceived difficulty of a step and the desire for AI assistance is not always consistent. For instance, although ``\textit{Understand Contribution Guidelines}'' was not perceived as overly difficult, participants expressed a significant need for assistance in this step, while the step ``\textit{Coding a Solution}'' presented the opposite situation.
\end{mdframed}

\vspace{-0.2cm}
\subsection{RQ3: Newcomers' Perceptions of the Design Strategies}
In RQ3, our objective is to validate our interpretation of the design strategies proposed by the participants in Phase I. To achieve this goal, we developed a prototype based on these design strategies and gathered participants' perceptions of it. Understanding the perceptions of the OSS newcomers who encounter challenges during their onboarding process is essential for evaluating the suitability of the proposed solutions.

\subsubsection{Implemented Prototype}
\label{sec:Implemented_Prototype}
\begin{figure}[htbp]
    \centering
    \begin{minipage}{0.42\textwidth}
        \centering
        \includegraphics[width=\textwidth]{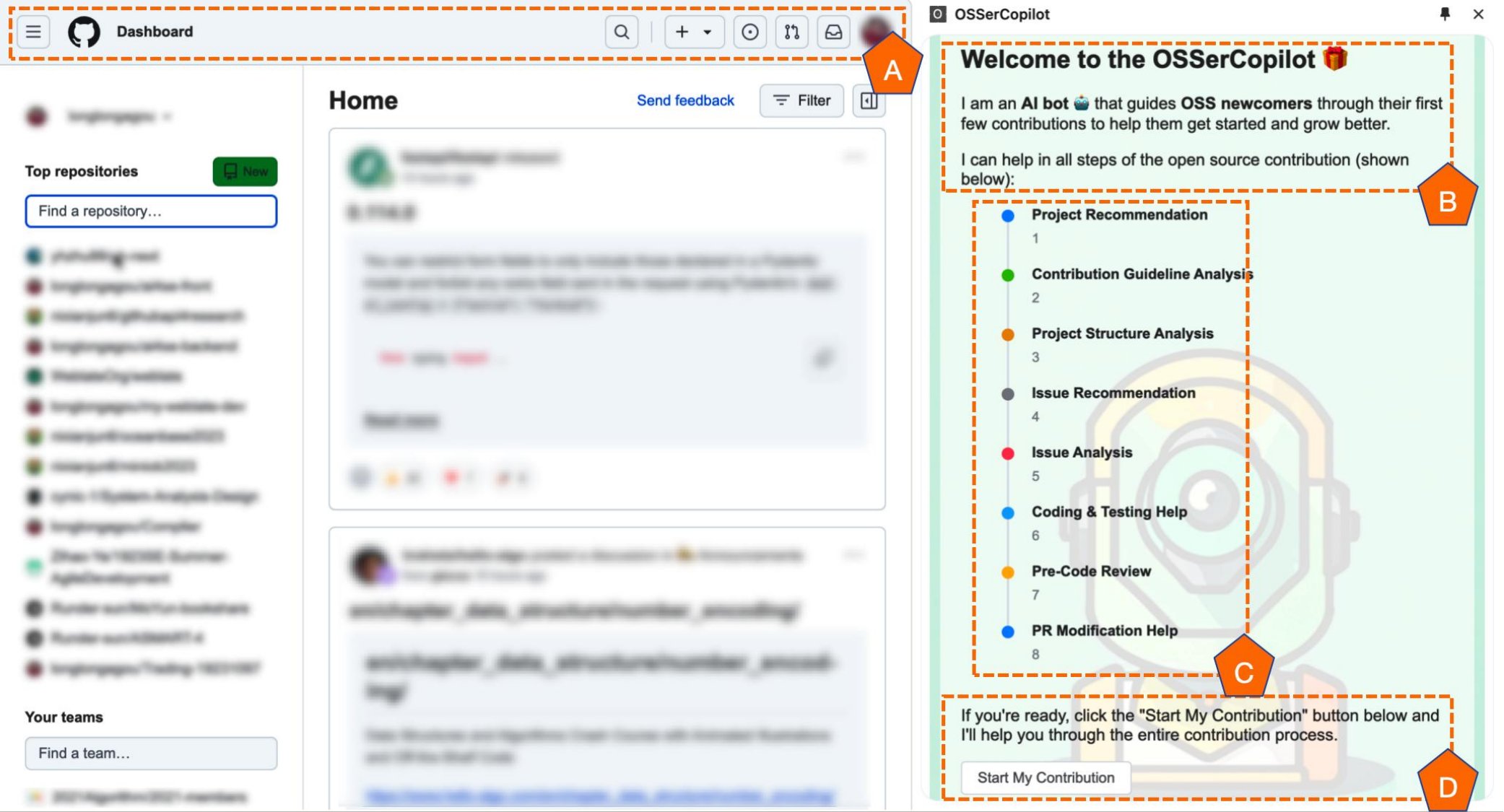}
        \vspace{-0.3cm}
        \caption{Homepage of Prototype}
        \vspace{-0.3cm}
        \label{fig:prototype_homepage}
    \end{minipage}
    \hspace{0.001\textwidth}
    \begin{minipage}{0.55\textwidth}
        \centering
        \includegraphics[width=\textwidth]{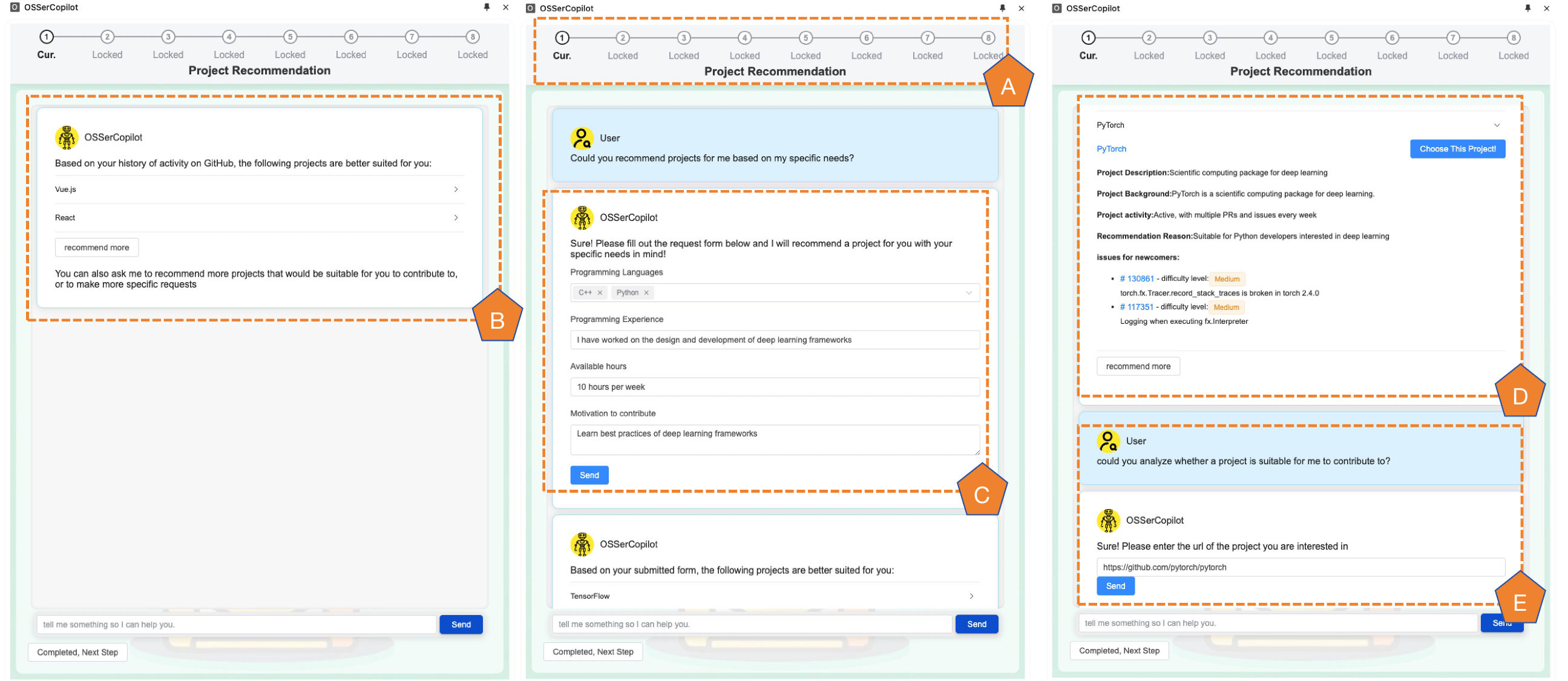}
        \vspace{-0.3cm}
        \caption{Recommend Project Page of Prototype}
        \vspace{-0.3cm}
         \label{fig:prototype_RecProj}
    \end{minipage}
\end{figure}

We implemented our prototype as a Web plugin, named ``\textit{OSSerCopliot}'', serving newcomers with sidebars as shown in Fig.~\ref{fig:prototype_homepage}. In accordance with the most participants' expectations, we integrated \textit{OSSerCopliot} with the most representative OSS development platform GitHub (see Fig.~\ref{fig:prototype_homepage}-A). Additionally, the homepage includes an introduction to \textit{OSSerCopliot} (see Fig.~\ref{fig:prototype_homepage}-B), an overview of its features (see Fig.~\ref{fig:prototype_homepage}-C), and provides an entry point to begin contributing (see Fig.~\ref{fig:prototype_homepage}-D). When newcomers click the ``Start My Contribution'' button, \textit{OSSerCopliot} will guide them through the contribution process and provide a progress bar above page helping them understand their current stage (see Fig.~\ref{fig:prototype_RecProj}-A). Due to space limitations, in the following, we use the step ``\textit{Discover an Interested Project}'' as a representative example to illustrate how we mapped the strategies onto the designed interface. The complete prototype, along with functional demonstration videos, is available in the supplemental material.

As shown in Fig.~\ref{fig:prototype_RecProj}, we implemented three approaches to assist newcomers in finding an interested project, corresponding to the first three design strategies for this step (see Table~\ref{tab:design_strategies}). First, when newcomers access the project recommend interface for the first time, \textit{OSSerCopliot} automatically recommends suitable projects based on their activity history on OSS platform (see Fig.~\ref{fig:prototype_RecProj}-B). Newcomers can also request that \textit{OSSerCopilot} recommend projects based on their specific needs, for which \textit{OSSerCopliot} will provide a customized form to gather newcomers' detailed and unique requirements, such as programming language, technical frameworks, and learning goals (see Fig.~\ref{fig:prototype_RecProj}-C). Finally, if the newcomer has a specific project in mind, they can provide the project URL, and \textit{OSSerCopliot} will analyze its suitability for contribution based on relevant project data (see Fig.~\ref{fig:prototype_RecProj}-E). Additionally, when recommending projects, \textit{OSSerCopliot} will also provide specific details, including reasons for recommendation, activity level of the project, and the set of GFIs (see Fig.~\ref{fig:prototype_RecProj}-D), corresponding to the fourth design strategy in this step.

\subsubsection{Overall Perception}

Participants provided feedback on \textit{OSSerCopilot} through interviews and a follow-up survey, which assessed their satisfaction with \textit{OSSerCopilot} at various stages of the onboarding process. Similar to Section~\ref{sec:Expectations_Ranking_at_Different_Steps} in RQ2, we compared participants' expectations for AI mentor assistance with their satisfaction with our prototype at each step.  However, in RQ3, we used average Likert scale ratings (from [1: Strongly Disagree] to [5: Strongly Agree]) to assess whether the prototype (presenting our implementation of design strategies) met newcomers' expectations, focusing on satisfaction rather than identifying the most challenging step. The results, shown in Fig.~\ref{fig:expectation_implementation}, indicate that participants were generally positive about the prototype. The radar chart of satisfaction scores overlaps with expectation scores, suggesting that \textbf{the prototype met participants' expectations}. Participants praised \textit{OSSerCopilot} for its clear, step-by-step guidance. P3 said, ``\textit{the lack of clarity in the contribution process made my work inefficient. The AI mentor helps by showing me which stage I'm at and what to do next}''. P6 noted, ``\textit{finding a suitable project to contribute to is challenging. If the prototype were realized, it would make this step easier}''.

\subsubsection{Perceived Usefulness, Ease of Use, and Potential Future Use}

In this section, we present the TAM survey results on the prototype's perceived usefulness, ease of use, and potential future use. We conducted a reliability analysis using Cronbach's Alpha~\cite{bland1997statistics} to ensure internal validity and consistency. The Alpha values were 0.87 for usefulness and 0.82 for ease of use, both above the 0.70 threshold, indicating high reliability~\cite{carmines1979reliability}.

\begin{figure}[htbp]
    \centering
    \begin{minipage}{0.35\textwidth}
        \centering
        \includegraphics[width=\textwidth]{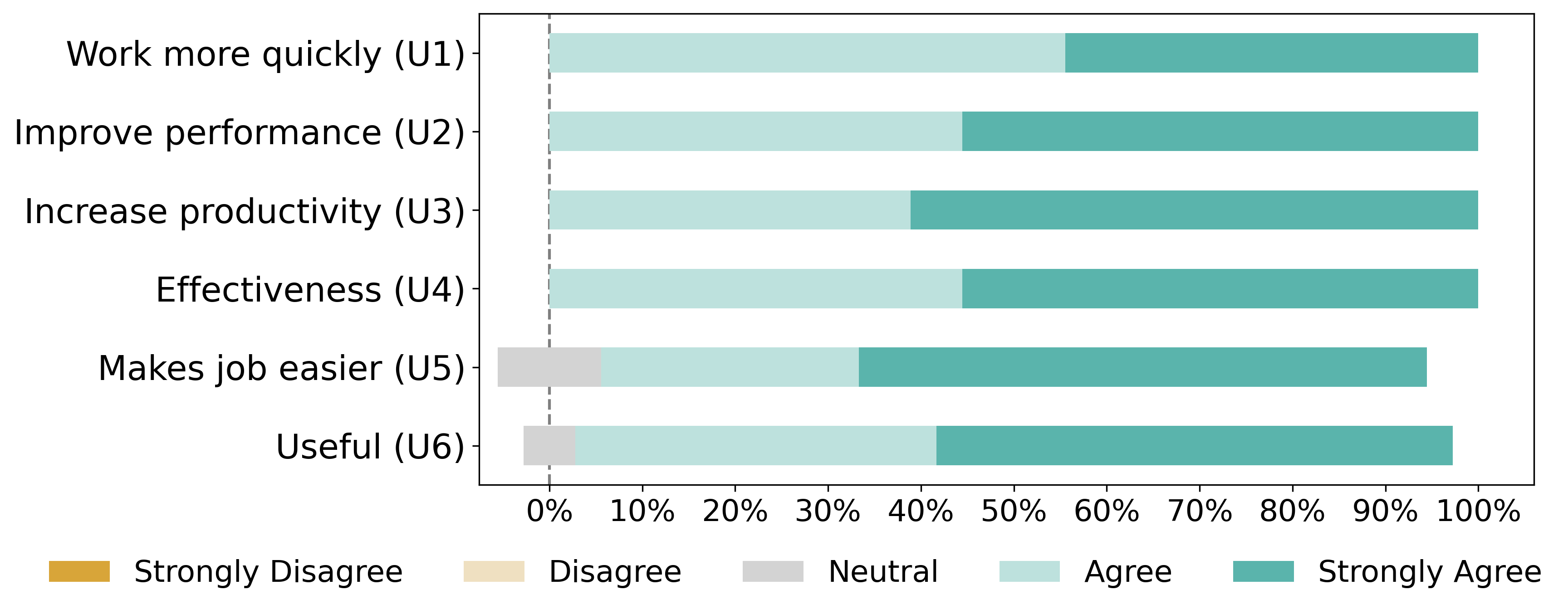}
        \caption{Perceived Usefulness}
        \label{fig:tam_usefulness}
    \end{minipage}
    \hspace{0.001\textwidth}
    \begin{minipage}{0.35\textwidth}
        \centering
        \includegraphics[width=\textwidth]{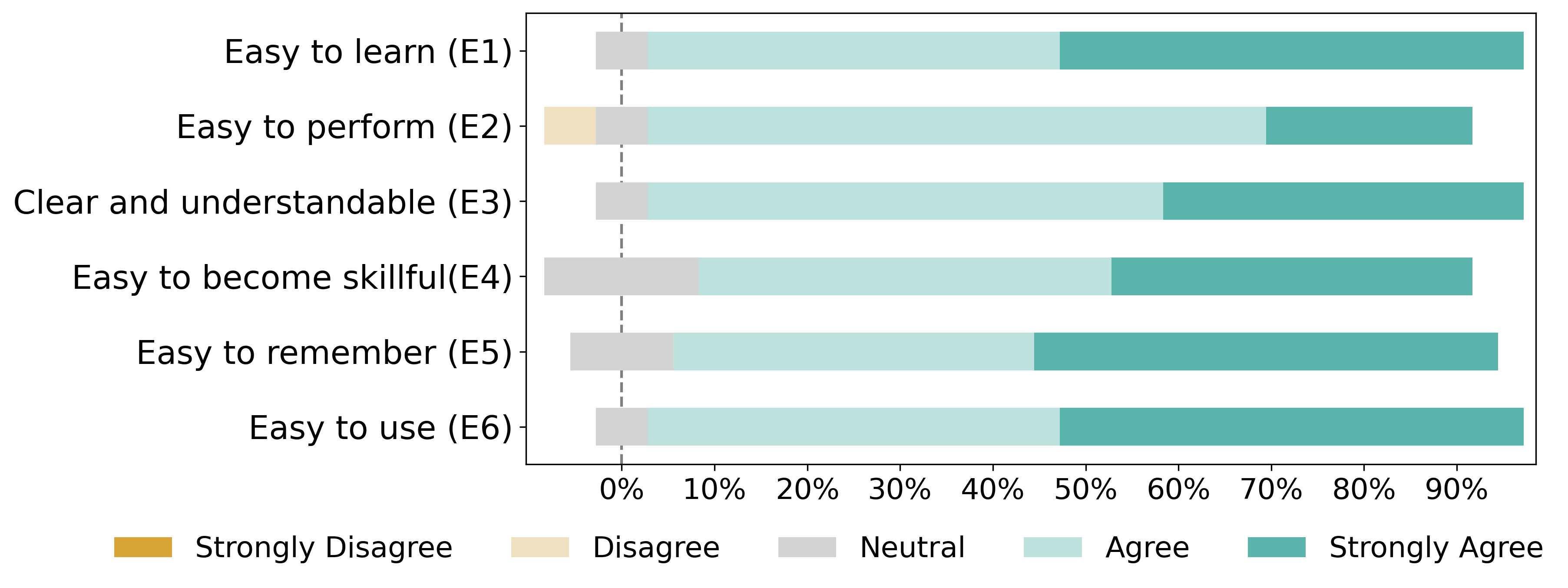}
        \caption{Perceived Ease of Use}
         \label{fig:tam_easeofuse}
    \end{minipage}
    \hspace{0.001\textwidth}
    \begin{minipage}{0.27\textwidth}
        \centering
        \includegraphics[width=\textwidth]{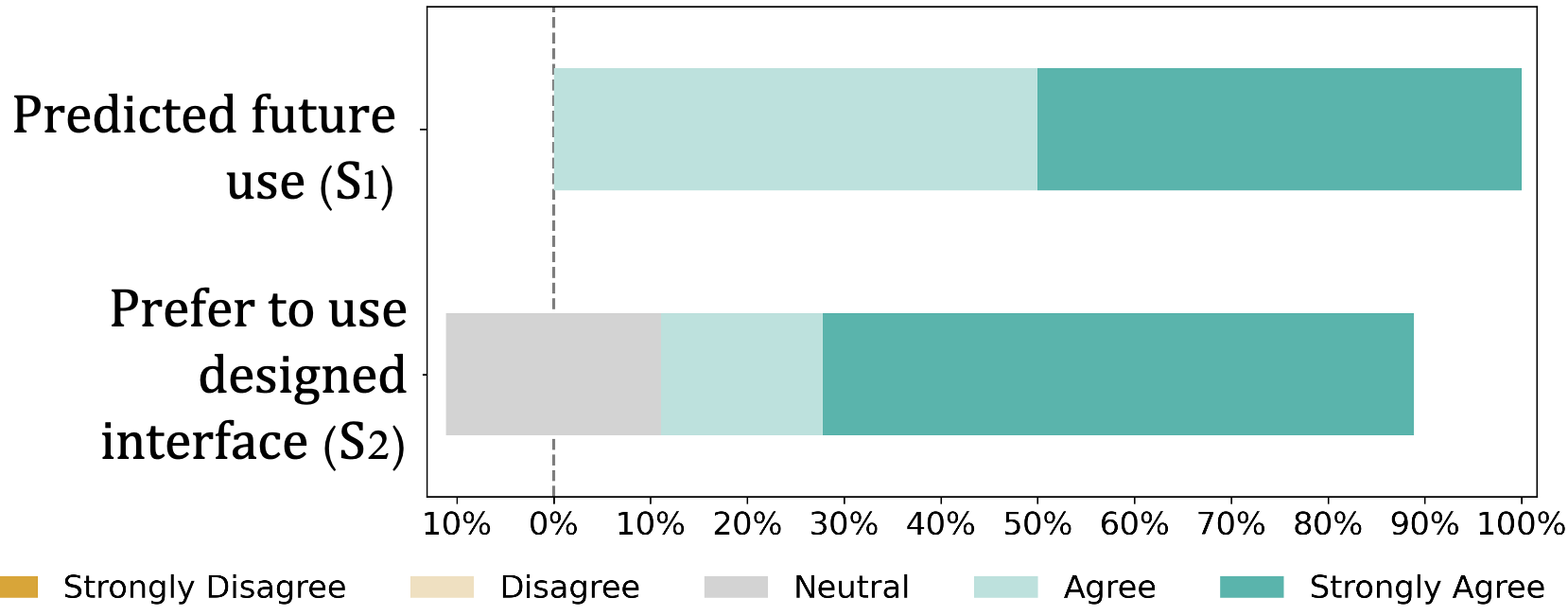}
        \caption{Self-predicted Future Use}
         \label{fig:tam_spfu}
    \end{minipage}
\end{figure}

\textbf{Usefulness of \textit{OSSerCopilot}.} Most participants found \textit{OSSerCopilot} useful. The detailed results are presented in Fig.~\ref{fig:tam_usefulness}. Notably, none of the participants disagreed with any item related to the usefulness of the \textit{OSSerCopilot}—all items received over 88\% agreement or strong agreement. Specifically, all participants (strongly) agreed that using \textit{OSSerCopilot} helps them \textit{work more quickly} (U1), \textit{improve performance} (U2), \textit{increase productivity} (U3), and \textit{increase effectiveness} (U4).

\textbf{Ease of Use of \textit{OSSerCopilot}.} Fig.~\ref{fig:tam_easeofuse} presents the distribution of responses for ease of use items, with over 83\% of participants agreeing or strongly agreeing. Only one participant expressed disagreement regarding the \textit{ease of using \textit{OSSerCopilot} to perform their tasks} (E2). Our suggestions for improving \textit{OSSerCopilot} in Section~\ref{sec:Suggested_Improvements}, are likely to positively impact its ease of use.

\textbf{Self-predicted Future Use.} Fig.~\ref{fig:tam_spfu} reports participants' self-predicted future use of \textit{OSSerCopilot}. We observe that all participants agreed or strongly agreed that they will use \textit{OSSerCopilot} if it is available in the future (S1). In comparison to the current approach for supporting newcomers in the onboarding process, 78\% of participants (strongly) agreed that they preferred \textit{OSSerCopilot}.

\subsubsection{Suggested Improvements}
\label{sec:Suggested_Improvements}

Participants also suggested 8 improvements for \textit{OSSerCopilot}, as shown in Table~\ref{tab:improvements}. These include feature enhancements and new functionalities. For example, in the ``\textit{Grasp Project's Structure}'' step, they proposed ``\textit{visualizing relationships between modules with flowcharts or class diagrams}'' (8) to aid understanding. For issue recommendations, participants wanted to ``\textit{suggest similar or related issues after one is resolved}'' (3) to improve contribution continuity. They also suggested ``\textit{mapping GFIs to a tree diagram of the project's structure}'' (2) and ``\textit{guiding newcomers to review their knowledge after resolving an issue}'' (3). UI improvements such as ``\textit{syntax highlighting and personalized settings}'' (6) were recommended for better usability.

\vspace{0.2cm}
\begin{mdframed}
[linecolor = gray!100,linewidth = 3pt,
innerleftmargin = 3pt, topline=false, rightline=false, bottomline=false, leftline=true, innerrightmargin = 3pt,innertopmargin = 3pt, innerbottommargin = 3pt,backgroundcolor = gray!30]
\textbf{Summary for RQ3:} We implemented an AI mentor prototype named ``\textit{OSSerCopilot}'', based on the elicited design strategies. Overall, the prototype effectively met participants' expectations. Participants perceived ``\textit{OSSerCopilot}'' as an useful and easy to use AI mentor and expressed a willingness to use it in the future, indicating the suitability of the design strategies. We also identified eight potential improvements for our designed prototype, which further refines the AI mentor's design strategies.
\end{mdframed}

\begin{table}[]
\centering
\scriptsize
\caption{Suggested Improvements to the Prototype (\# = Number of participants who mentioned this item)}
\label{tab:improvements}
\renewcommand{\arraystretch}{1.1} 
\begin{tabular}{l|c}
\hline
\rowcolor[gray]{0.8} \textbf{Suggested Improvements} & \textbf{\#} \\
\hline
Visualize the relationships between modules or functions within the project codebase through flowcharts or class diagrams. & 8 \\
\hline
UI-related improvements such as syntax highlighting in code blocks. & 6 \\
\hline
Recommend similar or related issues after newcomers resolve an issue. & 3 \\
\hline
After newcomers complete an issue, guide newcomers to review the knowledge gained throughout the process. & 3 \\
\hline
Present a ``Contributions in Progress'' list on homepage, which outlines the current status and entries for each contribution. & 2 \\
\hline
Provide automatic project deployment. & 2 \\
\hline
Map GFIs onto the tree diagram of project's structure when recommending issues. & 2 \\
\hline
Integrate with IDE. & 1 \\
\hline

\end{tabular}
\end{table}

\subsection{RQ4: Gaps between Existing Research and Design Strategies}
In RQ4, we conducted a literature review to identify the gaps between our design strategies and current research. After manual analysis, we finally identified 537 papers in total from high-quality conferences and journals in SE and AI. Table~\ref{tab:number_of_articles} presents the average ranking scores of participants' expectations for AI mentor's assistance at each step, alongside the number of papers we retrieved that align with the design strategies for each step. During our analysis, we observed that existing AI support for the last two steps was similar, and participants' average ranking scores for their expectations at both steps were close (2.68 and 2.53 respectively). Thus, in Table~\ref{tab:number_of_articles} and the following discussion, we have combined these two steps for a unified analysis.

\begin{table}[htbp]
\centering
\scriptsize
\caption{Average Ranking Scores of Participants' Expectations and the Number of Papers at Each Step}
\renewcommand{\arraystretch}{1} 
\begin{tabular}{lcc|lcc}
\toprule
\multicolumn{1}{c}{Step} & \#Expectation & \#Paper & \multicolumn{1}{c}{Step} & \#Expectation & \#Paper\\
\midrule
 1. Discover an Interested Project & 7.21 & 1 & 5. Analyze the Issue & 4.26 & 25 \\
 2. Understand Contribution Guidelines& 4.74 & 1 & 6. Code a Solution & 4.53 & 231 \\
 3. Grasp Project Structure & 6.53 & 5 & 7. Test Implementation & 3.68 & 129 \\
 4. Identify a Task & 5.53 & 8 & 8. Make the First Contribution  & 2.61 & 137 \\
 & & & \ \ \ \& Modify the PR & & \\
\bottomrule
\end{tabular}
\begin{tablenotes}
\footnotesize         
    \item[1] \#Expectation = Average Ranking scores of participants' expectations for AI mentor's assistance at each step.
    \item[2] \#Paper =  Number of papers retrieved that may satisfy design strategies of this step.
\end{tablenotes}
\label{tab:number_of_articles}
\end{table}

As shown in Table~\ref{tab:number_of_articles}, there are a great number of papers corresponding to the last four steps, indicating that existing research provides more robust design strategy support for these steps. For instance, papers related to ``\textit{Code a Solution}'' encompass a range of tasks, including code generation, code completion, code translation, and code style alignment, addressing the three design strategies (see Table~\ref{tab:design_strategies}) proposed by participants at this step. A total of 129 papers related to testing also align with the design strategies of ``\textit{test units generation}'' and ``\textit{test cases generation}'' in ``\textit{Test Implementation}''. Among the 137 papers related to PR submission or modification, 13.9\% focus on commit message generation, 16.8\% on code review, and the remaining 69.3\% on code repair, partially addressing the needs of this step (i.e., the first and third strategies in step ``\textit{Make the First Contribution}'' and the first strategy in step ``\textit{Modify the PR}''). 

Compared to last four steps, there is significantly less research related to the earlier steps, despite participants' high expectations for AI mentor's assistance during these stages. While several studies focus on automatic techniques for ``README'' files, we found only one paper~\cite{10.1145/3611643.3616288} related to the step ``\textit{Understand Contributing Guidelines}''. In this paper, the authors proposed a machine learning classifier to automatically categorize paragraphs in CONTRIBUTING files, aiding newcomers better understand the guidelines. A similar situation was observed in step ``\textit{Discover an Interested Project}'', which ranked highest in participants' expectations. Only one paper~\cite{10.1145/3382494.3410690} explored leveraging machine learning to recommend suitable topic tags for OSS projects on GitHub, which may facilitate project discovery for newcomers. 

Although there has been increasing research on how to use AI to automatically resolve GitHub issues~\cite{yang2024swe, tao2024magis, xia2024agentless}, newcomers' needs of step ``\textit{Analyze the Issue}'' have not yet been fully met. Among 25 papers corresponding to this step, 23 studies employed techniques like IRBL (Information Retrieval-Based Bug Localization), which helps developers locate defect-related code by analyzing the natural language descriptions (such as bug reports or issue descriptions), relating to design strategy of ``\textit{navigate and locate relevant code of the issue}'' in this step. However, these studies did not address the code localization for other types of issues (e.g., new feature issues). The remaining two papers focused on bug reproduction, satisfying our design strategy of ``\textit{reproduce the issue}''.

When filtering papers related to Step ``\textit{Grasp Project Structure}'', we specifically focused on studies that examine project-level code structure, rather than those centered on method-level or function-level analysis. We finally obtained five papers. Four of these studies are related to the first improvement strategies participants proposed in Phase II (see Table~\ref{tab:improvements}). For example, Husák et al.~\cite{10174103} provided a tool called SLICITO, capable of analyzing and visualizing a C\# program structure. Licker et al.~\cite{8812082} automatically uncover information between the files and modules of a project. Chi et al.~\cite{10.1145/3173574.3174029} and Chen et al.~\cite{8812043} linked the development UI with its corresponding code, helping developers understand unfamiliar source code. The remaining paper~\cite{9463018} directly studied the source code summarization at the project level.

As discussed in the related work, several studies have explored how AI technologies can assist project maintainers in labeling issues as GFIs or in recommending GFIs to newcomers. The eight papers filtered at step ``\textit{Identify a Task}'' also focus on these two topics. Notably, these studies only addressed the second design strategy for this step, indicating that research on recommending issues to newcomers is still insufficient.

\vspace{0.2cm}
\begin{mdframed}
[linecolor = gray!100,linewidth = 3pt,
innerleftmargin = 3pt, topline=false, rightline=false, bottomline=false, leftline=true, innerrightmargin = 3pt,innertopmargin = 3pt, innerbottommargin = 3pt,backgroundcolor = gray!30]
\textbf{Summary for RQ4:} 
Through the literature review, we identified discrepancies between the current research trend on AI and our proposed design strategies. Specifically, research is relatively scarce in the areas where newcomers highly anticipate AI mentor assistance, such as ``\textit{Discover an Interested Project}'' and ``\textit{Grasp Project Structure}'', indicating that significant gaps remain in the research on leveraging AI to support newcomers integrate and grow. 
\end{mdframed}


\section{Threats to Validity}
We discuss the potential threats to our findings and how we addressed or mitigated them.

\textbf{Generalizability of the Findings.} We validated our design strategies with a prototype as a GitHub plugin because most participants preferred AI mentor integrated with existing platforms, and GitHub is the most representative. However, this means that our findings may not apply to other OSS platforms or contexts. Although we do not anticipate big differences in these platforms, future research is necessary to test these strategies on different platforms to improve generalizability.

\textbf{Information Saturation.} 
We continued recruiting participants and conducting interviews until we determined that no new significant information was emerging. In line with the guidance of Strauss and Corbin~\cite{strauss1997grounded}, data collection can be concluded once it is sufficiently comprehensive and new insights are no longer being uncovered. To gain a comprehensive understanding of the onboarding process for OSS newcomers, particularly their needs regarding AI mentors, we conducted interviews with a diverse group of 19 participants, a number consistent with and even more than similar studies~\cite{rezwana2023user, wong2021using, wu2019fair}. This group, selected to ensure diversity in gender and background, included individuals from an OSS development course as well as newcomers (students and software developers) from various OSS communities. This approach provided a well-rounded perspective on the challenges and needs associated with AI mentoring.

\textbf{Collection of Relevant Literature.}
In RQ4, we conducted a literature review, which may introduce several threats, including selection bias, incomplete keyword coverage, time frame limitations, subjectivity in manual filtering, and the potential exclusion of important papers due to a limited number retrieved. To mitigate these risks, we employed several strategies: First, we restricted our search to high-quality journals and conferences to ensure the relevance and rigor of the selected studies, reducing selection bias. We also refined the keyword list with input from a professor specializing in automated software engineering, improving the accuracy of our keyword searches. Additionally, by reviewing the top 100 most relevant papers for each keyword, we balanced the need for comprehensiveness with the practical constraints of manual analysis. Finally, discussions between the first two authors helped minimize subjectivity in the filtering process.

\textbf{Reliability of results.} To mitigate potential threats to the interpretation of qualitative data such as researcher bias, we have taken several validation steps. First, to direct data collected, we carefully designed a 5-minute animated video and guided participants through contributing scenarios as a starting point for thinking about the future and constantly reminded them that they were not constrained by current technological limitations. Next, we employed the constant comparison method, continuously comparing each emerging interpretation with existing findings as it emerged from the qualitative analysis. Regular discussions were held to refine the coding process. We also developed a prototype based on the extracted design strategies and gathered feedback from developers through TAM to verify their accuracy. Finally, we performed a reliability check on the TAM questionnaire items to ensure the robustness of the TAM instrument. These validation steps reduce the impact of bias and coding inconsistencies on the overall findings, thereby improving the reliability of our findings.

\section{Conclusion}
In this paper, we took the first step toward AI-assisted newcomer integration in OSS communities by analyzing current onboarding processes, eliciting AI mentor design strategies, and identifying research gaps. We found that newcomers' current onboarding mainly relied on search engines and community resources, with little direct community help. They face significant challenges in step ``\textit{Grasp project's structure}'', ``\textit{Discover an interested project}'', and ``\textit{Identify a task}''. By applying Participatory Design Fiction,  we identified 32 design strategies for AI mentor. Participants also found the prototype ``\textit{OSSerCopilot}'' we developed based on the strategies both useful and user-friendly, expressing a willingness to use it in the future, demonstrating the relevance and effectiveness of the proposed strategies. Our literature review revealed research gaps in areas where newcomers highly anticipate AI assistance. This research has the potential to revolutionize the current newcomer-expert mentorship and provides valuable insights for OSS project maintainers, researchers and tool designers aiming to develop and enhance AI mentor systems.

\textbf{For OSS project maintainers,} it is suggested to provide more newcomer-friendly project information, such as clearer contribution process guidance, more detailed project structure descriptions, and concise documentation specifically tailored for newcomers. This documentation should emphasize essential information for beginners while condensing less relevant details. Additionally, they should carefully assess whether the project's GFIs are truly suitable for newcomers and consider using AI to label GFIs to avoid cognitive bias, thus improving efficiency and accuracy. By taking these steps, OSS communities can better support newcomers' integration and foster community growth. Our findings also highlight the significant potential of AI mentors, which are expected to greatly reduce the workload of experts in guiding newcomers. 

\textbf{For AI mentor designers,} our study provides rich insights into the specific challenges faced by newcomers during their onboarding, and the steps where participants most struggled with and desired AI assistance. This analysis helps designers identify the key challenges newcomers face and develop AI mentor focused on the areas where assistance is most needed. The 32 design strategies proposed by newcomers provide direct guidance for developing AI mentor in line with their needs and expectations. By considering these findings, AI mentor designers can make informed decisions to create effective and newcomer-friendly AI mentor.

\textbf{For SE researchers,} our literature review identified discrepancies between the
current research trend on AI and our proposed design strategies. Future research can focuses on areas where newcomers highly anticipate AI mentor assistance while corresponding research is relatively scarce, such as ``\textit{Discover an Interested Project}'' and ``\textit{Grasp Project Structure}''. Additionally, while emerging tools like GitHub Copilot Workspace show potential to address some our proposed design strategies, their effectiveness in OSS onboarding scenarios remains unclear. By exploring the potential of AI to help newcomers in these key stages, SE researchers can uncover valuable insights for better supporting  newcomers' integration in OSS development.

\section{Data Availability}
The supplementary materials, including the fiction story, video, prototype, and codebook used in this study, are available online: \url{https://figshare.com/s/dba3b22dd1b93bf1329b}.

\begin{acks}
This work is supported by the National Natural Science Foundation of China (Grant Nos. 62202022, 62272445, and 62332001), the Fundamental Research Funds for the Central Universities, and the exploratory elective projects of the State Key Laboratory of Complex and Critical Software Environments.
\end{acks}

\bibliographystyle{ACM-Reference-Format}
\bibliography{ref}

\end{document}